\documentclass[onecolumn,10pt]{asme2ej}

\usepackage{epsfig} %% for loading postscript figures

\usepackage{graphicx,color}  
\usepackage{bm}        % for math
\usepackage{amssymb}   % for math
\usepackage{amsmath}

\title{Role of kidney stones in renal pelvis flow}

\author{C. R. Constante-Amores
    \affiliation{
    Mathematical Institute \\ 
    University of Oxford \\
    Oxford OX2 6GG, UK \\
    Email: crc15@ic.ac.uk
    }
}

\author{L.~Kahouadji\\
    \affiliation{
        Department of Chemical Engineering\\
        Imperial College London\\
        London SW7 2AZ, UK\\
    }
}

\author{J.~G.~Williams\\
    \affiliation{
    Department of Urology and Pelvic Health\\ Boston Scientific Corporation\\
    Marlborough\\
    Massachusetts
    }
}

\author{B.~W.~Turney
    \affiliation{
    Nuffield Department of Surgical Sciences \\ 
    University of Oxford\\
    John Radcliffe Hospital\\
    Oxford OX2 6GG, UK \\
    }
}

\author{S.~Shin
    \affiliation{
    Dept. of Mech. and System Design Engineering \\ 
    Hongik University \\
    Seoul 04066, Republic of Korea \\
    }
}

\author{J.~Chergui
    \affiliation{
    Universit\'e Paris Saclay \\ 
    LISN\\
    CNRS \\
    91400 Orsay, France
    }
}

\author{D.~ Juric
    \affiliation{
    Universit\'e Paris Saclay \\ 
    LISN \\
    CNRS \\
    91400 Orsay, France;\\
    DAMTP\\
    University of Cambridge\\
    Cambridge CB3 0WA, UK\\
    }
}

\author{D.~E.~Moulton
    \affiliation{
    Mathematical Institute \\ 
    University of Oxford \\
    Oxford OX2 6GG, UK \\
    }
}

\author{S. L. Waters
    \affiliation{
    Mathematical Institute \\ 
    University of Oxford \\
    Oxford OX2 6GG, UK \\
    }
}

\begin{document}

\maketitle

%%%%%%%%%%%%%%%%%%%%%%%%%%%%%%%%%%%%%%%%%%%%%%%%%%%%%%%%%%%%%%%%%%%%%%
\begin{abstract}
\textcolor{black}{
Ureteroscopy is a commonly performed medical procedure to treat stones in the kidney and ureter using a ureteroscope. Throughout the procedure, saline is irrigated through the scope to aid visibility and washout debris from stone fragmentation. The key challenge that this research addresses is to build a fundamental understanding of the interaction between the kidney stones/stone fragments and the flow dynamics in the renal pelvis flow. }
We examine the time-dependent flow dynamics inside an idealised renal pelvis in the context of a surgical procedure for kidney stone removal. %, extending previous  work  by \cite{williams_2020,williams_2021}, who showed  how vortical flow structures can  hinder mass transport in a canonical two-dimensional domain.
 Here, we examine  the time-dependent evolution of these vortical flow structures in three-dimensions, and incorporate the presence of rigid kidney stones. 
We perform direct numerical simulations, solving the transient Navier-Stokes equations in a spherical domain. 
Our numerical predictions for the flow dynamics in the absence of stones are validated with \textcolor{black}{available} experimental and numerical data, %from \cite{williams_2020}, 
and the governing parameters and flow regimes are chosen carefully in order to satisfy several clinical constraints. 
The results  shed light on the crucial role of flow circulation in the renal cavity and its effect on the trajectories of rigid stones. We demonstrate that stones can either be washed out of the cavity along with the fluid, or be trapped in the cavity via their interaction with vortical flow structures. Additionally, we study the effect of multiple stones in the flow field within the cavity in terms of the kinetic energy, entrapped fluid volume, and the clearance rate of a passive  tracer  modelled via an advection--diffusion equation. We demonstrate that the flow in the presence of stones features a higher vorticity production within the cavity compared with the stone-free cases. 
\end{abstract}

%%%%%%%%%%%%%%%%%%%%%%%%%%%%%%%%%%%%%%%%%%%%%%%%%%%%%%%%%%%%%%%%%%%%%%

%%%%%%%%%%%%%%%%%%%%%%%%%%%%%%%%%%%%%%%%%%%%%%%%%%%%%%%%%%%%%%%%%%%%%%

\section{Introduction}
The renal pelvis is a funnel-like cavity inside the kidney which connects directly to the ureter, leading to the urinary bladder. The kidney's primary function -- to remove waste and excess fluid from the body -- may be disrupted by the presence of  kidney stones in the renal pelvis, resulting in potentially life-threatening conditions. Kidney stones affect up to $10\%$ of the  global population, and thus constitute a significant healthcare concern \cite{Romero2010,Cortes2011,Turney_2012}. 
\textcolor{black}{
Chemically, the majority ($80-85\%$) of kidney stones are composed of calcium oxalate and are often due to poor diet/lifestyle and/or dehydration from low fluid intake. There are many other types of kidney stones including calcium phosphate, urate, infection stones and metabolic stones (e.g. cystine)}. 
%Chemically, kidney stones are composed of calcium, phosphate, or other constituents of foods, and are usually due to poor lifestyle and/or dehydration from low fluid intake. 
Very small stones, the size of a grain of sand, may form and pass without causing any symptoms but larger stones which grow to more than $5$ mm \cite{Khan_2016} can become stuck in the renal pelvis. \textcolor{black}{Some stones grow to fill the kidney (staghorn stones) and can measure several centimeters across.}  Enlarged stones may cause blockages within the urinary system, often resulting in severe patient discomfort  \cite{Segura_1985,Miller468}. Thus, there is a clear necessity for a clinical procedure to efficiently remove stones from the renal pelvis. 

One method for stone removal, \textit{ureteroscopy},  uses a laser to ablate and fragment the stone to reduce its size and facilitate natural evacuation through the ureter. The laser is inserted through the central lumen of the ureteroscope. To further assist with stone clearance, a \textcolor{black}{flow} of saline is continuously introduced to the renal pelvis through this central lumen, in a process called \textit{irrigation}. The fluid exits the renal cavity via a narrow gap between the ureteroscope and surrounding ureter. This return flow is often facilitated by the presence of an \textit{access sheath}, a hollow tube surrounding the scope shaft, which increases the gap -- now between the scope shaft and access sheath wall -- via which saline (and stone fragments) can exit. Figure \ref{fig_intro}a provides a schematic representation of the placement of the ureteroscope and access sheath during ureteroscopy. 

Prior to laser-induced fragmentation, kidney stone sizes range  between $5$ and $10$ mm ($66\%$), and  $10$ and $15$ mm ($28\%$) \cite{Cui2013}. After ablation, the stone is reduced to a large number of fragments, ranging in radii from $R_s = 1$ $\mu$m to $R_s = 1$ mm. The stone particle sizes therefore vary by several orders of magnitude, and particle-fluid interaction plays a non-trivial role in the renal-flow dynamics and associated kidney stone removal. Different particle-fluid interaction behaviours are anticipated depending  on the size of the stone fragments: 
(i) for small stone sizes ($R_s\lesssim 200~\mu$m),  stones behave as a passive tracer without affecting the flow dynamics of the renal cavity,
(ii) for large stone sizes ($R_s\gtrsim 5000~\mu$m),  stones will settle in the cavity under gravity, and
(iii) for intermediate stone sizes ($200~\mu\rm m \lesssim R_s \lesssim5000~\mu$m), a fully-coupled interaction between the rigid solid and the flow is expected \textcolor{black}{\cite{Keller_2021}}. We aim to develop a holistic understanding of the different fluid-structure interaction mechanisms at play, including the multi-scale effects of varying solid particle sizes and their interactions with a viscous fluid. An ability to model stone fragment behaviour -- and its dependence on the number, size, and arrangement of stone particles in the renal pelvis -- has the potential to advise clinical procedure and device design to optimise the efficiency of ureteroscopic stone removal. 

Previous work by \cite{lykoudis_1970} and
\cite{Yang11932} have proposed theoretical models of ureteral fluid mechanics, considering the ureteroscope as a single conduit. However, during the ureteroscopic procedure, there is both the forward flow into the cavity from the scope lumen, as well as the return flow from the kidney to the ureter through the access sheath. 
\cite{oratis_2018} incorporated the effect of the return flow through the access sheath with a lumped-parameter model to relate flow rate, kidney pressure, and ureteroscope and access sheath geometry, modelling the kidney as a linearly compliant material with  constant stiffness.
\cite{Williams_pressure} also considered a lumped-parameter model incorporating a more biologically accurate exponential constitutive law for the kidney compliance, and studied the role of auxiliary `working tools' -- e.g. laser fibres -- which are passed through the central lumen of the ureteroscope and affect resistance to saline flow.

These lumped parameter models were able to predict the renal pressure due to irrigation and its dependence on flow rate and scope/access sheath geometry, but the precise nature of the flow within the renal pelvis was not considered.  More recently, \cite{williams_2020} modelled the renal pelvis in an idealised two-dimensional, rectangular geometry, and studied in detail the steady flow patterns and their subsequent effect on the clearance time of a passive tracer, using a combination of numerical methods and high-speed imaging techniques. 
Williams \textit{et al} revealed the intricate vortex structures in a two-dimensional cavity, and
demonstrated the connection between the  wash-out time (defined as the time required for $90\%$ of the initial tracer to leave the cavity) and the vortex characteristics within the cavity, concluding that large vortices combined with low tracer diffusivity lead to prolonged wash-out times.
In a follow-up study, \cite{williams_2021} studied the role of the inflow/outflow channel geometries on  the wash-out time of a passive tracer. Using shape optimisation techniques with an objective function based on properties of the steady flow field, Williams \textit{et al} demonstrated that
changing the ureteroscope shape results in a reduction of  the size of vortical flow structures, which in turn leads to a decrease in wash-out times.
Although these studies provide valuable insights into the relationship between the underlying flow dynamics and the washout time of a passive tracer, this knowledge is only transferable to ureteroscopy in the regime where stone fragments are extremely small, and thus unable to influence flow characteristics. In reality, larger stone pieces are present during ureteroscopy; thus it is necessary to consider the two-way coupling of flow and stone dynamics for an accurate understanding of stone removal. Additionally, there is a need to study full three-dimensional flows in cavities  to unravel the flow physics and the associated  stones trajectories.

In this study, for the first time, we interrogate the two-way coupling between the fluid flow within the kidney and the transport of kidney stones, and determine the behaviour of stone wash-out during ureteroscopy.   Considering  the stones as rigid, non-porous solid objects inside a three-dimensional idealised spherical cavity representing the renal pelvis that is connected to coaxial cylinders representing the ureteroscope and  sheath, we perform time-dependent three-dimensional direct numerical simulations using a fictitious domain method with direct forcing approach 
to resolve the interaction between the rigid solids and the flow. 
To orientate our study, the values of the governing parameters are chosen to be consistent with clinically-realisable values.

The paper is organised as follows: Section 2 presents the governing equations, numerical set-up and the validation of the numerical method. 
Section 3 \textcolor{black}{presents } the results which are focused on fluid-structure interaction in both the absence and the presence of stones. Finally, \textcolor{black}{a discussion of the results }, and concluding remarks and future perspective work are summarised in Section 4 \textcolor{black}{and 5, respectively}.

\section{\textcolor{black}{Methods}}

\begin{figure}
\centering
\includegraphics[width=1\linewidth]{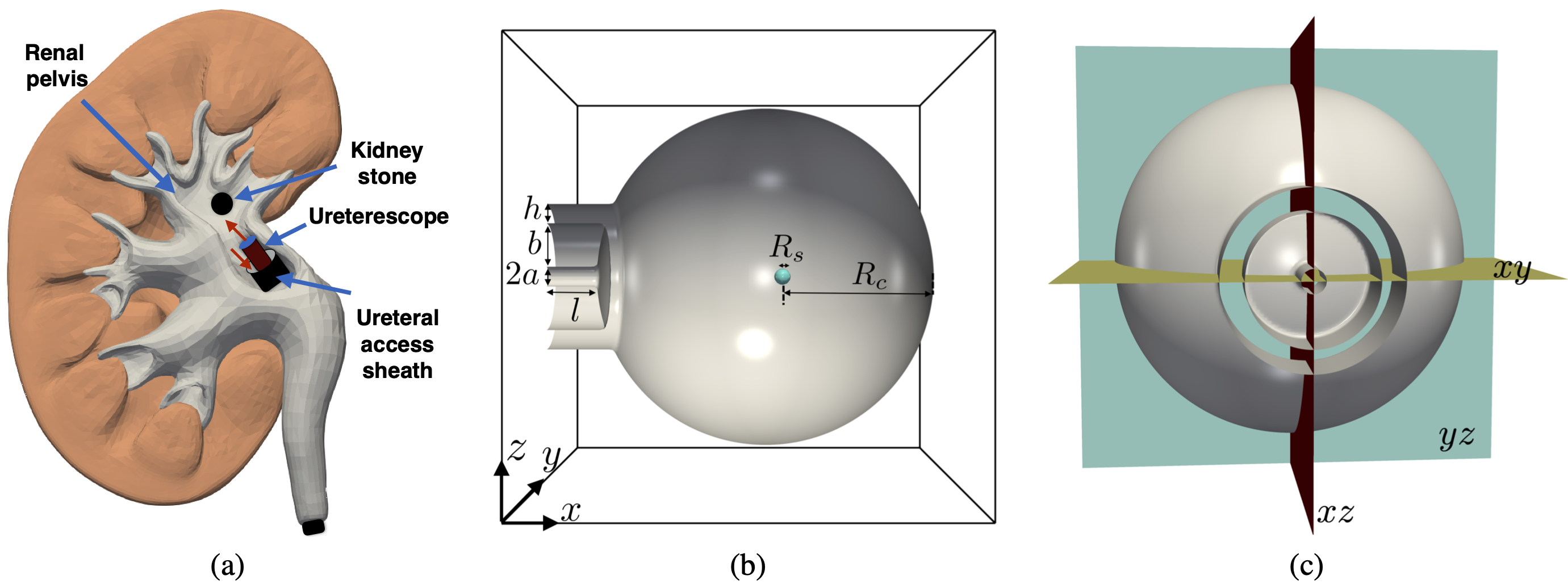}  
\caption{(a) Schematic representation of the regions of fluid flow during the ureteroscopic procedure; we highlight the ureteroscope via which the flow enters the renal pelvis, and a ureteral access sheath in which the irrigation flow exits the cavity. The flow direction inside of the renal pelvis is shown with red arrows. (b)
Three-dimensional representation of the idealised geometry of the renal cavity considered in this study with a single rigid spherical stone at center. The image shows a transversal cut to aid visualisation of the spherical stone, it also  highlights the nondimensional parameters of the cavity geometry in a three-dimensional Cartesian domain, $\textbf{x} = (x,y,z)$ which is discretised for our numerical computation by a $(256)^3$ uniform mesh. The center of the spherical cavity corresponds to $\textbf{x}=(0,0,0)$. \textcolor{black}{Red and blue arrows show the inlet and outlet channels, respectively}. (c) Three-dimensional representation of the  $xy$-, $xz$- and $zy$-planes that will be used throughout  the discussion of the results.
\label{fig_intro}}
\end{figure}

With the aim of studying the flow dynamics of kidney stone removal, we perform direct numerical simulations of the transient, incompressible Navier-Stokes equations in a three-dimensional domain focusing, for this present study, on the case of the absence of gravity. 
We assume spherical rigid non-porous stones which move freely in the computational domain owing to their interaction with the cavity flow. 
Figure \ref{fig_intro}b shows a schematic representation of the cavity considered in this study.
To simplify the tremendous complexity of the renal cavity, we consider a \textcolor{black}{stiff, non-deformable} sphere of radius $R_c$, which is connected to a  ureteroscope and a ureteral access sheath.
The ureteroscope and  sheath are considered to be coaxial cylinders, in which flow enters the cavity via the inner scope channel of radius $a$ and exits the cavity through the gap (width $h$) between the access sheath and scope shaft of radius $b$.
Renal stones are characterised by an irregular shape, but for the sake of simplicity, we have considered  them as perfect spherical non-porous rigid solids of radius $R_s$. 
They are placed initially in the centre of the cavity, if not explicitly stated otherwise.

Following the immersed boundary approach \cite{Peskin_jcp_1977}, we use a one-fluid formulation to couple the fluid-structure interaction (FSI) problem.
Within this formulation, parameters $\rho_l$ and $\mu_l$ are the density and viscosity of the fluid, respectively, and $\rho_s$ is the density of the solid.  

We consider dimensionless variables
\begin{equation} \label{scales}
\quad \tilde{\mathbf{x}}=\frac{\mathbf{x}}{a},
\quad \tilde{t}=\frac{t}{a/U}, 
\quad \tilde{\textbf{u}}=\frac{\textbf{u}}{U},
\quad \tilde{p}=\frac{p}{\mu_l U/a}, 
\end{equation}
\noindent 
where t, $\textbf{u}$, and $p$ represent time, velocity, and pressure, respectively, and tildes denote dimensionless quantities.  We nondimensionalise spatial coordinates $\mathbf{x} = (x,y,z)$ with respect to the radius of the scope lumen $a$, $\mathbf{u} = (u,v,w)$ with respect to the average irrigation flow velocity $U$, and time with respect to the timescale of the flow, $a/U$. The dimensionless time-dependent, incompressible Navier-Stokes equations for a Newtonian viscous fluid are thus

\begin{equation}\label{div}
 \nabla \cdot \tilde{\textbf{u}}=0,
\end{equation}

\begin{equation}
\label{NS_Eq}
\tilde{\rho} \left ( \frac{\partial \tilde{\textbf{u}}}{\partial \tilde{t}}+\tilde{\textbf{u}} \cdot\nabla \tilde{\textbf{u}} \right) + \nabla \tilde{p}  = 
\frac{1}{Re}  \nabla^2 \tilde{\textbf{u}}
%\nabla\cdot  \left [  (\nabla \tilde{\textbf{u}} +\nabla \tilde{\textbf{u}}^T) \right ]
,
\end{equation}
where 
$\tilde{\rho}=\rho_s/\rho_l + \left(1 -\rho_s/\rho_l\right)
\mathcal{H}\left( \tilde{\textbf{x}},\tilde{t}\right)$ 
%and we have a similar expression for the viscosity field, $\tilde\mu$, 
where $\mathcal{H}\left( \tilde{\textbf{x}},\tilde{t}\right)$ is zero in the solid stones and unity in the fluid. %Additionally, $\tilde\mu=1$ in the fluid.
\noindent
%
%The density field is  $\tilde{\rho}=\rho_s/\rho_l + \left(1 -\rho_s/\rho_l\right)\mathcal{H}\left( \tilde{\textbf{x}},\tilde{t}\right)$, where $\rho_s$ is the density of the solid stones, and we consider a similar expression for the viscosity field, $\tilde\mu$.  Following the immersed boundary approach \cite{Peskin_jcp_1977}  we use a  smoothed Heaviside function, $\mathcal{H}\left( \tilde{\textbf{x}},\tilde{t}\right)$ which is zero in the solid stones and unity in the fluid. 
The dimensionless  Reynolds number, $Re =\rho_l U a /\mu_l$, in equation \ref{NS_Eq} 
relates the inertial to viscous forces.
All the variables appearing in the equations and boundary conditions are rendered dimensionless using the aforementioned scalings, unless stated otherwise. 

A no-slip condition  for the immersed solid-fluid  boundary, $\Gamma$, is enforced, taking into account rigid body motion due to translation and rotation of the stones

\begin{equation}
\label{solidBC}
{\tilde{\textbf{u}}}_\Gamma=\tilde{\textbf{V}}_s + \tilde{ \Omega}  \times  {\tilde{\textbf r}}_c 
\end{equation}

\noindent
where ${\tilde{\bf V}}_s$ represents the translation velocity of the solid (non-dimensionalised with respect to $U$), ${\tilde{ \Omega}}$ stands for its angular velocity vector (non-dimensionalised with respect to $U/a$) and ${\tilde{\bf r}}_c$ the position vector of the centroid of the solid (non-dimensionalised with respect to $a$). ${\tilde{\bf V}}_s$ and ${\tilde{\Omega}}$ can be computed by averaging over the solid region. \textcolor{black}{ For the collision modeling we use the the well-known impulse-response model (for further  information we refer the reader to \cite {baraff1997introduction}).}

\subsection{Initial and domain boundary conditions}
Henceforth we drop tildes, and all quantities are now dimensionless. The simulations are initialised with fluid and stones at rest in the absence of gravity. Solutions are sought subject to Poiseuille inlet flow.  Thus a parabolic profile for the inflow channel is specified (i.e., $u=2 (1- r^2)$), where $r=\sqrt{\left(y-y_o\right)^2 + \left(z-z_o\right)^2}$,  where $y_o$ and $z_o$ are the (dimensionless) coordinates of the centre of the inflow channel.
As outflow conditions, we impose Neumann boundary conditions for the velocity in the streamwise direction (i.e., perpendicular to the outflow face), and the spanwise velocity components are zero.  A Dirichlet boundary condition is imposed for the pressure.  This condition is a result of imposition of zero normal stress at the outlet together with the assumption of fully-developed streamwise flow.
Additionally, we impose no-slip boundary conditions on the cavity walls.

\subsection{Numerical method and validation}

The Navier-Stokes equations are solved using classic finite volume techniques applied on a uniform staggered grid \cite{Harlow_pf_1965}. A multigrid iterative method is used for solving the elliptic pressure Poisson equation that arises in the projection method \cite{Chorin_mc_1968} when enforcing the incompressibility condition (Equation \ref{div}). With respect to the spatial derivatives, standard centered-difference discretisations are used, except for the nonlinear term, which makes use of a second-order essentially nonoscillatory (ENO) scheme \cite{Shu_1989,Sussman_1994}.
The Lagrangian motion of the solid centroid (i.e., $\rm d \textbf{r}_c/ \rm d \textbf{t} = \textbf{V}_s$) is computed using a second-order in time Runga-Kutta algorithm.  The boundary condition Eq. \ref{solidBC} is enforced using the Direct Forcing approach of \cite{Fadlun} where the desired value of velocity is imposed directly on the solid/fluid boundary.  In addition to the extensive analysis and validation provided in \cite{Fadlun} we validate our own implementation of the Direct Forcing method with results shown in the Appendix.
In this section we have provided only a brief synopsis of the numerical approach, however a detailed description of the immersed boundary, fictitious domain and direct forcing methods implemented here can be found in  \cite{Shin_ijnmf_2009,Shin_2020,Fadlun,Glowinski}. \textcolor{black}{ The code is wholly written by the authors in Fortran 2008 and uses a domain decomposition strategy for parallelization with MPI.}

The validity of the numerical method was benchmarked against the experimental and numerical work of \cite{williams_2020} for a two-dimensional cavity with excellent agreement (more information in the Appendix).
In terms of the mesh characteristics, the computational domain is composed of
a cubic uniform Cartesian grid, and the inlet extends in the $x-$direction. With respect to the resolution, we have ensured that the presented results are mesh-independent, and therefore for a uniform mesh resolution of $(256)^3$ (i.e., $a/\Delta {\bf x}=6.4$, where $\Delta {\bf x}$ stands for the cell-size), the results do not significantly change with decreasing cell size (see Appendix for more details).  
Additionally,  extensive mesh studies for turbulent two-phase jets  and surface-tension-driven phenomena using the same numerical method can be found in 
\cite{Constante_gfm,constante_jets}.

\subsection{Physical parameters}

The chosen flow parameters for this study are consistent with values used during ureteroscopy. In the clinic, irrigation flow rates range from $50$ mL/min to $200$ mL/min, although introduction of a laser fibre (for stone fragmentation) to the scope channel may  reduce the flow rate by about half.  
The injected fluid is considered to be water, with constant physical properties (i.e., $\rho_l= 1000$ kg/m$^3$ and $\mu_l=10^{-3}$ Pa.s). Therefore, the Reynolds number lies in the range $220 < \text{Re} < 2000$. 

In the ureteroscopy procedure, the stone is first  fragmented by the laser to enable expulsion through the ureter. We consider  fragmented stone sizes ranging between  $200\mu \hbox{m}<R_s<1000 \mu $ m where there will be a fully-coupled interaction between the flow and stone trajectory. We take the stone density as $\rho_s=1900$ kg/m$^3$ \textcolor{black}{\cite{Zhong_1993}}. Gravity will play a minor role in the flow dynamics,  as indicated by the typical Froude number $Fr=U /\sqrt{g a} \sim O(10^2)$ (where $g$ stands for the gravitational constant); thus, the effect of gravity is not considered in this study.

The values for the cavity dimensions are identical to the  previous work performed by \cite{williams_2021}, and consequently, $a = 6 \times 10^{-4}$ m, $b = 2.5701 \times 10^{-3}$ m, and $h = 1.1598 \times 10^{-3}$ m. The size of the cavity was chosen to be $R_c=10^{-2}$ m, which could be seen as a representation of a small region of the renal pelvis.
The non-dimensional length of the inlet channel is set to $5$. 
\textcolor{black}{
We note that the values for $b$ and $h$ are slightly larger than the typical range of scope and access sheath dimensions for ureteroscopy procedures. These were chosen to agree with the values used in previous physical and numerical experiments of flow in rectangular domains, simulating the dynamics of ureteroscopic flow in the renal pelvis \cite{williams_2020,williams_2021}. This agreement allows us to investigate the effect of scaling up from two to three dimensions, and although exact flow solutions will depend on the values of $b$ and $h$, we anticipate reported trends in flow structure and the influence of flow on stone dynamics will be similar for smaller values of $b$ and $h$. Values for $b$ and $h$ more representative of ureteroscope and access sheath dimensions were chosen for the simulations in the Discussion (see figure \ref{perspectives}), demonstrating the potential for higher fidelity ureteroscopy simulations using the techniques outlined in this manuscript.
}

\section{Results}

In this section we first consider the effect of Reynolds number on three-dimensional flow in the absence of stones, before subsequently analysing the effects of both single and multiple stones. It is worth noting that after $t = 5000$, for the lowest Reynolds number ($\text{Re} = 50$), no further changes in the shape of the  streamlines or coherent structures were observed; however, for the higher Reynolds numbers ($\text{Re} = 250$, $500$), flow evolution continued and a steady-state was not reached.

\subsection{Three-dimensional flow dynamics in the absence of rigid stones \label{sec:no_stones}}

\begin{figure}
\centering
\includegraphics[width=1\linewidth]{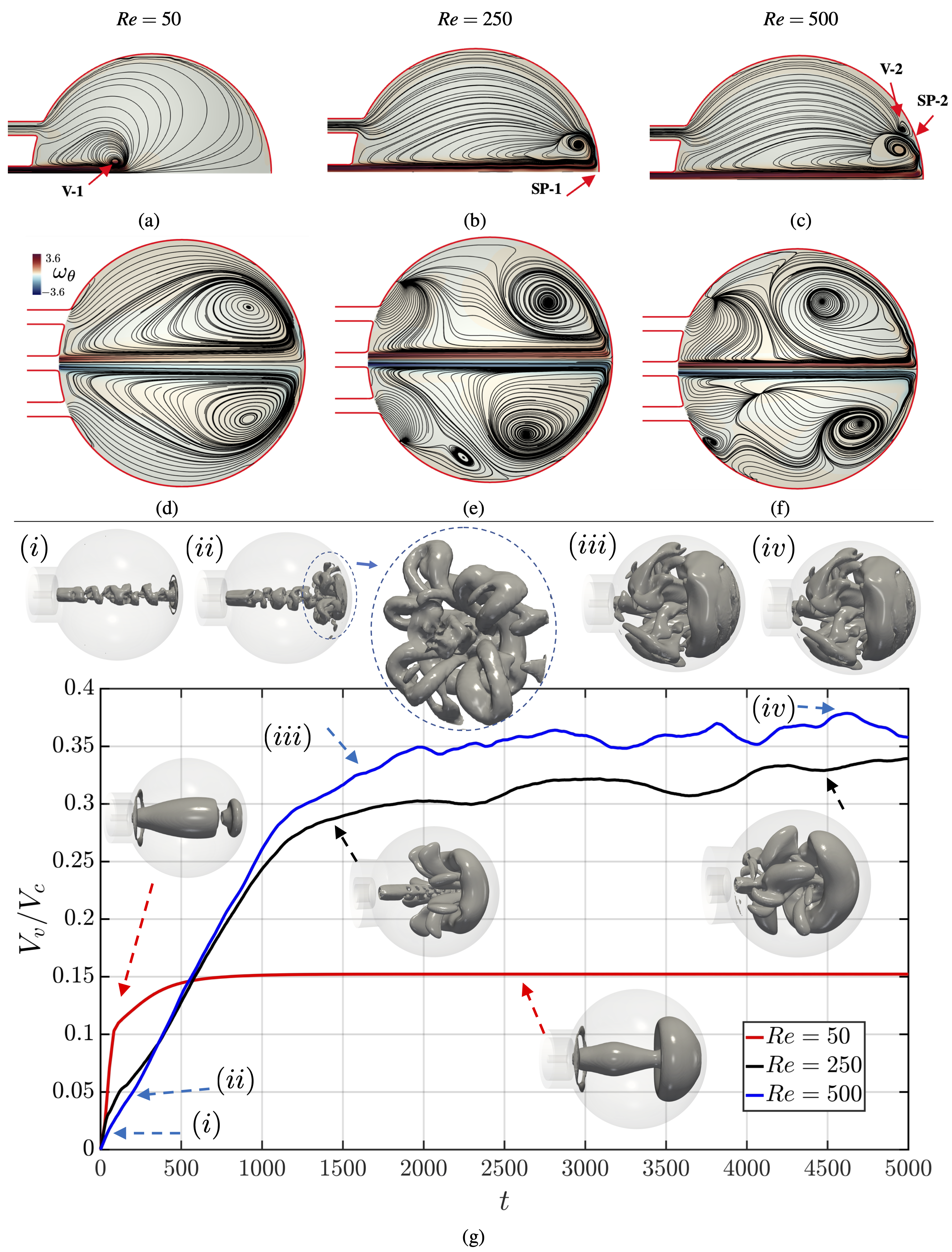}  
\caption{\label{Re_figure}
Time-dependent flow dynamics inside of the idealised renal cavity for $\text{Re}=(50,250,500)$. Panels (a)-(c) correspond to early flow dynamics for $Re=500$ at $\text{t}=(27,138,194)$ through two-dimensional representations of the predicted streamlines in a half of the $xy-$plane.
Panels (d)-(f) correspond to  two-dimensional representations of the predicted streamlines in a $xy-$plane together with the azimuthal vorticity $\omega_{\theta}$ at $t=5000$ for $\text{Re}=(50,250,500)$ in columns one to three, respectively;
Panel (g) shows the spatio-temporal evolution of the three-dimensional coherent structures by representing the entrapped fluid volume when $Q>0$, $V_v$ normalised by the volume of the cavity, $V_c$, for $\text{Re}=(50,250,500)$.
Additionally, we refer the reader  to Supplementary Material `Animation-Fig2.avi' to see a video from panel (e). }
\end{figure}

First, we focus on the transient dynamics of the jet.
Panels (a)-(c) of Figure \ref{Re_figure} show the instantaneous streamlines at early times for $\text{Re} = 500$. For visualization purposes, we display a two-dimensional streamlines onto the $xy$-plane (for sake of space only the half of the plane is shown). At early times of the injection, the jet dynamics are characterised by an axisymmetric behaviour. The homogeneous jet develops a `primary-vortex' (labelled `{\bf V-1}' in figure \ref{Re_figure}a) as a result of  the detachment of the velocity boundary layer from the  inlet nozzle, and its consequent roll-up in the quiescent medium. A similar physical mechanism for the formation of the leading vortex has been reported previously by \cite{gharib_rambod_shariff_1998,marugan_2013,constante_jets}. The primary vortex grows over time owing to the shear-driven interaction between the high-velocity discharge of the fluid into the quiescent medium. The  primary-vortex   elongates as it moves downstream until it impacts against the surface of the cavity wall resulting in the outward radial spread of the head-vortex (see figure \ref{Re_figure}b). The formation of a stagnation point (labelled `{\bf SP-1}' in figure \ref{Re_figure}b) near the cavity wall resulting in flow separation  is observed as a result of the impact of the jet against the wall.
For the high Reynolds numbers cases, we observe that the radial  expansion of the jet moving  outward  leads to an
adverse pressure gradient until a point  in which the jet-induced momentum  can no longer overcome the frictional effects of the  cavity wall resulting in the formation of a stagnation point (labelled `{\bf SP-2}' in figure \ref{Re_figure}c) and flow separation. This phenomenon leads to the detachment of a secondary-vortex, labelled `{\bf V-2}' in figure \ref{Re_figure}c (in agreement with \cite{Garimella_1995,Sexton,Sivasamy,Lee}). As the secondary vortex grows over time, a mutual-induction of {\bf V-1} and {\bf V-2} is predicted (see panel (ii) of figure \ref{Re_figure}g): 
a velocity induction by the primary-vortex over the secondary-vortex causes {\bf V-2} to go through {\bf V-1}'s centre. 
This mutual-induction mechanism is observed during the entire flow dynamics at high $Re$ numbers.

Panels (d)-(f) of figure \ref{Re_figure} highlight the instantaneous streamlines at $t = 5000$ for $\text{Re} = 50$, $\text{Re} = 250$, and $\text{Re} = 500$, respectively. For visualization purposes, we have also shown a two-dimensional streamlines onto the $xy$-plane. The  azimuthal component of the vorticity field $\boldsymbol{\omega}= \triangledown \times \textbf{u}$ (e.g. the circumferential direction around the axis of the jet), is shown as a contour field underneath the streamlines.
For the smallest Reynolds number, the jet ejection results in an axisymmetric  flow  with no formation of secondary vortices. A closer inspection of the same figure  shows that the areas of closed streamlines (i.e., the main vortical structure) are surrounded by streamlines which show a direct path between the inflow and outflow channels with no recirculation regions; these outcomes are in agreement with \cite{williams_2020,williams_2021}.
Increasing the Reynolds number results in the breaking of the symmetry of the flow patterns, and subsequently,  enhances the formation of complex flow patterns inside of the cavity.  The primary vortex is no longer aligned axisymmetrically in the cavity  as displayed in figures \ref{Re_figure}(e)-(f).  At higher $Re$, there is a reduction in the number of  direct  paths from the inlet to the outlet channels owing to the increase of the inertia within the cavity. 
Additionally, by inspection of the azimuthal vorticity in the $xy$-plane, we observe that the highest vorticity generation coincides with the velocity boundary layer between the injected-jet and the surrounding fluid, which leads to strong tangential flow with respect to the initial quiescent medium. 
Attention is now turned to the value of vorticity  (red/blue contours in Figures \ref{Re_figure}d-f) which increases as the Reynolds number increases. For $Re=50$,  the azimuthal vorticity component dominates over its streamwise counterpart (not shown) because of the  low inertia; however, as the Reynolds number increases, the streamwise component becomes more prominent, becoming of the same order of magnitude with respect to its azimuthal counterpart, explaining the disruption of the axisymmetric behaviour of the coherent structures as explained below.

Figure \ref{Re_figure}g displays the time-dependent flow dynamics by plotting the volume ratio $V_v/V_c$ occupied by the vortical structures (i.e., recirculation regions $V_v$), predicted by the Q-criterion inside of the cavity (with volume $V_c$). The Q-criterion measures the dominance  of vorticity $\omega$ over that of strain $\bf{s}$, i.e., $Q = 1/2(|\boldsymbol{\omega}|^2 -|\bf{s}|^2)$ \cite{Hunt_CTR_1988}. 
The flow regions with a positive Q-criterion value are defined as the recirculation zone (i.e., vortex-dominated regions). Thus, the volume representation of figure \ref{Re_figure}g denotes flow regions in which $Q>0$. A similar approach was used by \cite{williams_2020,williams_2021} to quantify the vortex-regions for a two-dimensional system (i.e., det $\triangledown \textbf{u}>0$).

For $Re=50$, the  axisymmetric primary-vortex defines the global flow dynamics and the vortex-dominated region reaches a steady state resulting in a constant entrapped recirculating fluid within  the cavity (see  figure \ref{Re_figure}g). 
At higher $Re$, the Q-criterion predicts larger vortex-dominated regions inside of the cavity with a linear increase of the  entrapped-fluid volume at early times of the simulation (see figure \ref{Re_figure}g).
The increase in the rich dynamics of the flow structures is evident by observing the spatial development of the coherent structures (i.e.,  regions of dominant vorticity). 
At short times, the jet impacts the cavity wall (see panel (i) of figure \ref{Re_figure}g), and eventually results in the formation of secondary vortices which are advected towards the centre of the primary vortex (see panel (ii)) while the primary vortex grows over time. Once the primary vortex reaches its maximum size, the entire flow almost behaves as a steady state (see panels (iii)-(iv)).

To see the impact of the vortical structures on the trajectories of rigid stones, in the next section we turn our attention to the effect of flow dynamics on the  motion of stones within the cavity.

\subsection{Effect of flow dynamics on  rigid kidney stones }

\begin{figure}[h!]
\centering
\includegraphics[width=1\linewidth]{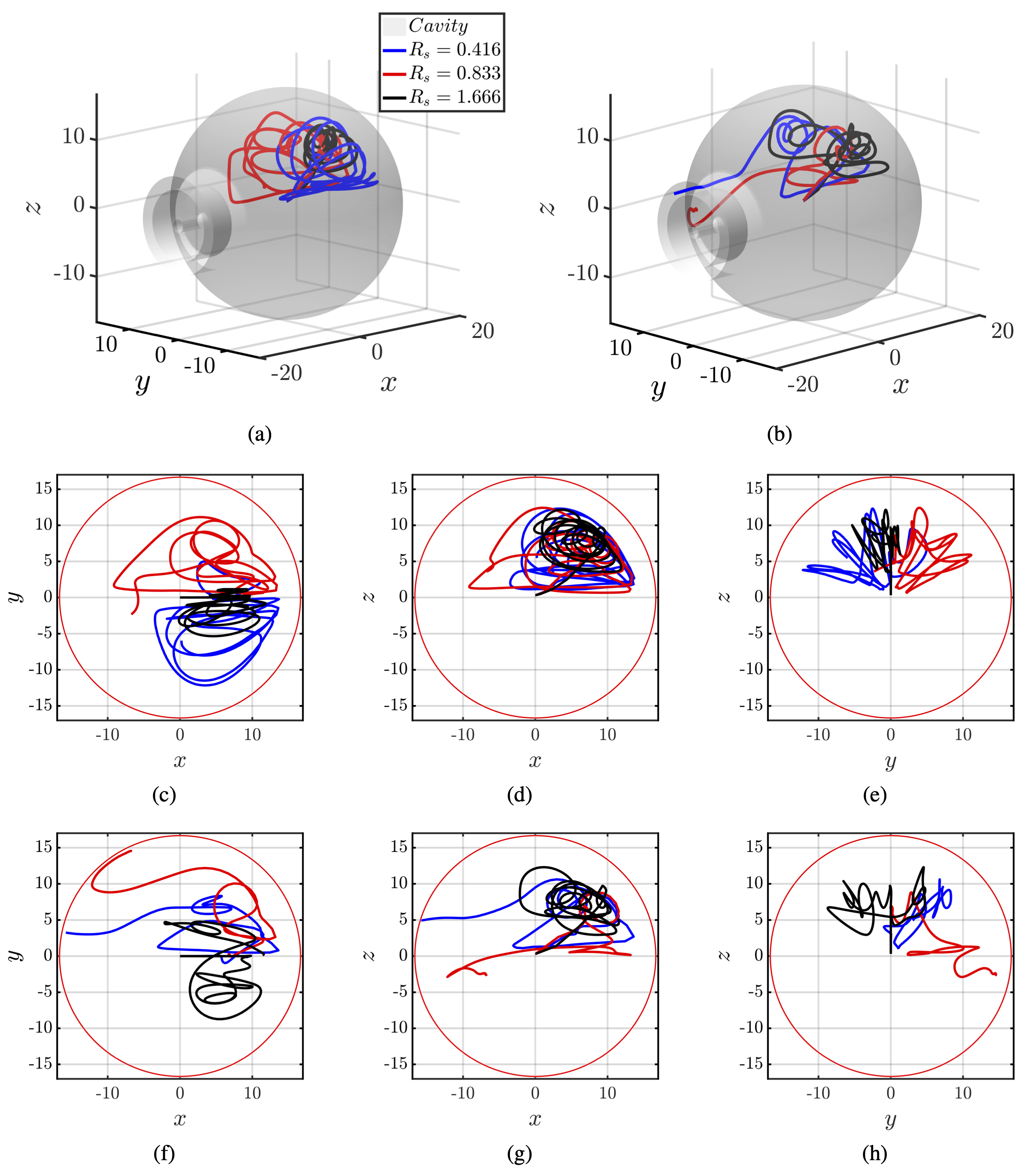}  
\caption{Trajectories of single rigid stones up to  $t=5000$. Panels (a) and (b) correspond to a three-dimensional representation of the trajectories for  $\text{Re}=250$ and $\text{Re}=500$, respectively. Panels (c)-(h) correspond to the projection of the trajectory onto the $xy$-, $xz$- and $yz$-plane for $\text{Re}=250$ (middle panels) and $\text{Re}=500$  (bottom panels), respectively. The red circles correspond to the limits of the cavity, where the center of the spherical cavity corresponds to $\textbf{x}=(0,0,0)$. \label{stone_trajectory} }
\end{figure}

\begin{figure}[hbt!] 
\centering
\includegraphics[width=1\linewidth]{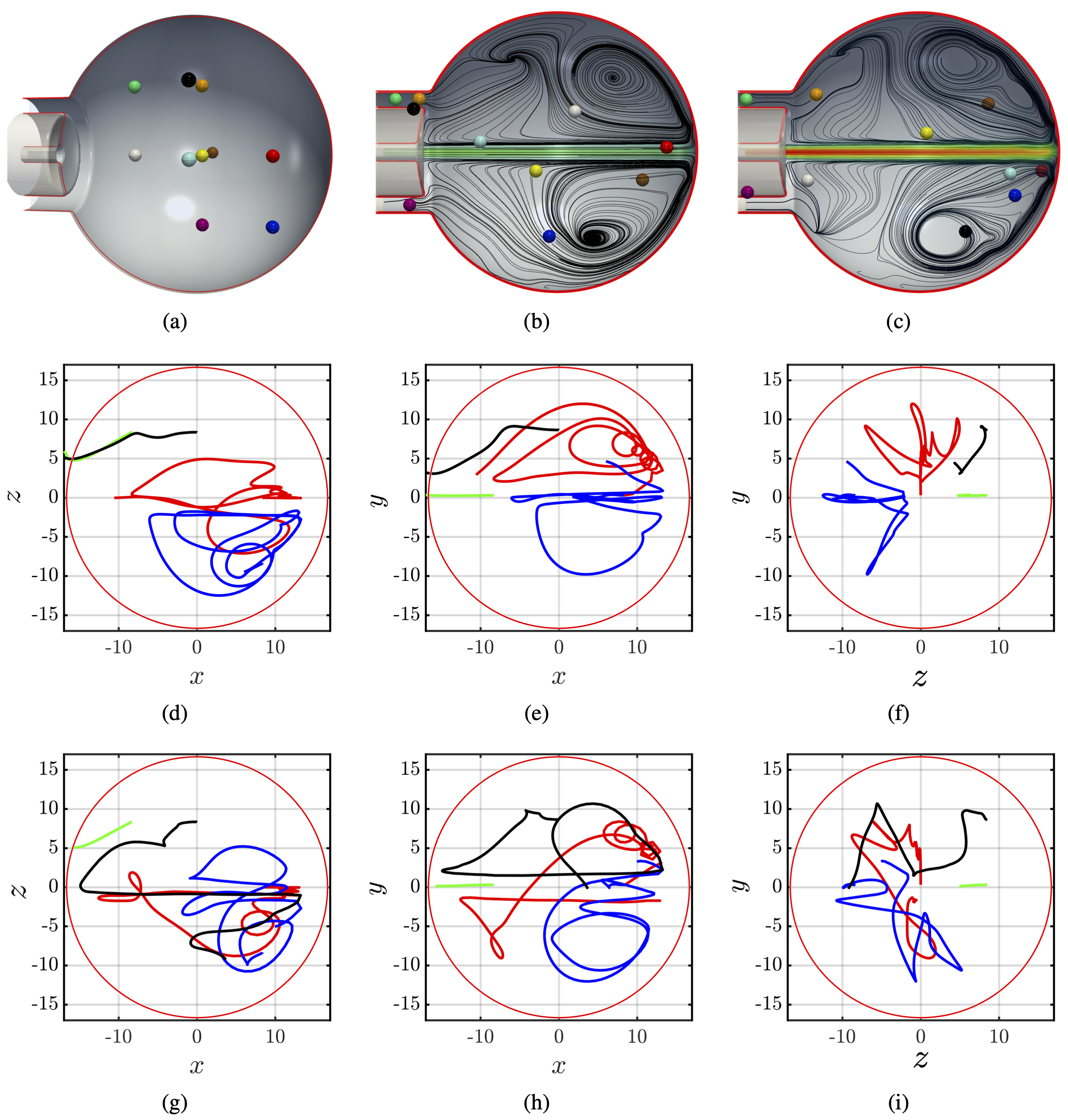}  
\caption{Trajectories of multiple rigid  stones (i.e., ten) of size $R_s=0.833$ up to $t=5000$. Panels (a)-(c) show the initial location of the stationary stones, the  location of the stones at $t=5000$ for  $\text{Re}=250$,  and $\text{Re}=500$, respectively. Additionally, a two-dimensional representation of the velocity streamlines in the xy-plane have been added to visualise the flow field.
Some stones have left the renal cavity and are found in the ureteral access sheath, meanwhile others are trapped in the primary vortex. 
Panels (d)-(i) correspond to the projection of the trajectory for four selected stones onto the $xz$-, $xy$- and $zy$-plane for $\text{Re}=250$ (middle panels) and $\text{Re}=500$  (bottom panels), respectively. The colour of the trajectories correspond to the coloured stones of panels (a)-(c). %The red circles correspond to the limit of the cavity, where the center of the spherical cavity corresponds to $\textbf{x}=(0,0,0)$
See Supplementary Material `Animation-Fig4.avi'.
\label{multiple} }
\end{figure}

In this section, we consider the effect of renal pelvis flows on the dynamics of a single rigid stone, of varying size, placed initially in the centre of the cavity (e.g., as displayed in Figure \ref{fig_intro}b). We vary the stone size between $0.416<R_s<1.66$ (nondimensional values). Additionally, we only report results for $\text{Re}=250$ and $\text{Re}=500$ as these correspond to larger values of inertia prompting richer flow dynamics within the cavity.

Figure \ref{stone_trajectory} shows  the trajectories of the stones over time by tracking their centroid as both three-dimensional trajectories (Figure \ref{stone_trajectory}a,b) and projections onto  the $xy$-, $xz$- and $zy$-planes (Figures \ref{stone_trajectory}c-h).   
At early stages of the simulation, the rigid stones are dragged towards the back of the cavity -- as a result of their interaction with the head-vortex -- and eventually collide against the cavity wall. 
The stones subsequently settle into a circulatory motion owing to their interaction with the local vorticity field.

For $\text{Re}=250$, the stones become trapped in the primary vortex, leading to their circulation in the vortex-dominated region (see Figure \ref{stone_trajectory}a).  Inspection of the stones' movement in each projection plane (Figure \ref{stone_trajectory}c-d) shows that the stones 
are  trapped in the primary vortex (we refer the reader to Supplementary material  `Animation-Fig3.avi' where we show  the transient dynamics of the stone of size $R_s=0.833$ together with the Q-criterion). Interestingly, the radii of the  trajectories grow over time, so it is expected that eventually the stones would be displaced to a non vortex-dominated region (e.g., near the cavity exit), where they will washout together with the fluid. Nonetheless, for the considered  non-dimensional time of $t<5000$, the rigid stones remain trapped in the primary vortex during the entire simulation.

Still considering $Re=250$, now we compare  the effect of stone size on the flow pattern and particle trajectory. For the smallest stone size  (e.g., $R_s=0.416$), the stone follows the primary vortex, and its circulation radius grows over time. 
The simulations predict  smaller circulation radii for the largest stone size and stronger entrapment in the vortical structure because the stones start stationary, so a larger stone would require a larger force to attain the same velocity.

For $\text{Re}=500$, richer flow dynamics are predicted owing to the  higher fluid inertia, which in turn leads to more complex stone trajectories (see Figures \ref{stone_trajectory}b,f-h). At early time in the simulations, the rigid stones similarly  enter into a circulatory motion due to their interaction with the primary vortex; however, the 
higher fluid inertia induces the stones to leave the primary vortical structure, which eventually results in the smallest stone, i.e., $R_s=0.416$, being flushed out of the cavity by the flow (at $t=3507$). Therefore, for the first time (to the best of the authors' knowledge), we have provided a simulation of stones being washed out of the cavity along with the fluid during the ureteroscopy procedure.

Above we have considered the effect of flow dynamics on single rigid stones placed in the centre of the cavity. To investigate how the initial position of the rigid stones affects the fate of the stone,  we consider ten stones within the cavity, initially placed  
 on random grid points as indicated in figure \ref{multiple}a. Panels of figure \ref{multiple}(b)-(i) show snapshots of the flow pattern and trajectory of stones over time via projections onto the  $xz$-, $xy$- and $zy$-planes.
We found that stones initially positioned near the centre of the cavity are 
pushed towards the outflow channel as the primary vortex grows in size, and eventually advect directly out of the cavity.
Conversely, stones initially placed further downstream of the injection point are entrapped in a circulatory motion owing to their interaction with the primary-vortex, and subsequently, the stone's dynamics are similar to those described above for a single stone that does not leave the domain. Therefore, some stones leave the cavity into the `ureteral access sheath', whereas others are still entrapped  in the main vortical structure (see Supplementary Material `Animation-Fig4.avi'). 

\textcolor{black}{
We acknowledge that to better understand the interaction between kidney stones and the flow cavity, it would be interesting to use Lagrangian coherent structures (LCS) based on particle trajectories to naturally find borderlines in the flow that partition different regions with different trajectory behavior. This will connect the particle initial and final locations. However, this is out of the scope of the manuscript, and should be addressed in futures studies.}

\subsection{Effect of rigid kidney stones on flow dynamics  \label{washout-time-section}}

In the previous section, we considered the effect of the flow field on the trajectories of small stones which are not likely to significantly affect the flow dynamics. In this section, we will consider the influence of larger kidney stones on the flow dynamics by exploring the fluid kinetic energy, the clearance rate of a passive tracer and the entrapped fluid volume when $Q>0$, depending on the number of stones, $N_p$.  
On this basis,
we consider a passive tracer of concentration $C(\textbf{x},t)$ within the cavity which is passively advected by the flow field. The  nondimensional advection-diffusion equation for the tracer concentration is expressed as  

\begin{equation} 
\frac{\partial C} {\partial t}+\textbf{u}\cdot \nabla C= \frac{1}{Pe} \nabla^2 C,
\end{equation}

\noindent
where $Pe= U a/\mathcal{D}$ represents the ratio of convective to diffusive time-scales (e.g., $\mathcal{D}$ stands for the tracer diffusion coefficient). Following \cite{williams_2021}, we assume that the  initial tracer concentration is evenly distributed  inside of the cavity, and zero elsewhere. We  have selected $Pe=500$, guided by the previous work from \cite{williams_2020}.
We assume no-flux boundary conditions on the walls of the cavity (i,e., $\partial C/ \partial \mathbf{n}=0$), and Neumann boundary conditions for $C$ on the outlet channels which allow the passive tracer to leave the computational domain. 
To quantify the effect of the rigid stones on the flow dynamics,  an approach proposed by \cite{williams_2020,williams_2021} was employed in this work which measures the clearance rate at  which the passive tracer leaves the cavity. Thus, the  nondimensional loss at time $t$ is expressed as

\begin{equation}
C_{\text loss}=\frac{\int_\mathcal{V}\left[C (\mathbf{x},0)-C(\mathbf{x},t)\right]d\mathcal{V}}{\int_\mathcal{V} \left[C(\mathbf{x},0)\right]d\mathcal{V}}
\end{equation}

\noindent
where $\mathcal{V}$ refers to the reduced volume of the cavity based on the number of stones.

\begin{figure}
\centering
\includegraphics[width=1\linewidth]{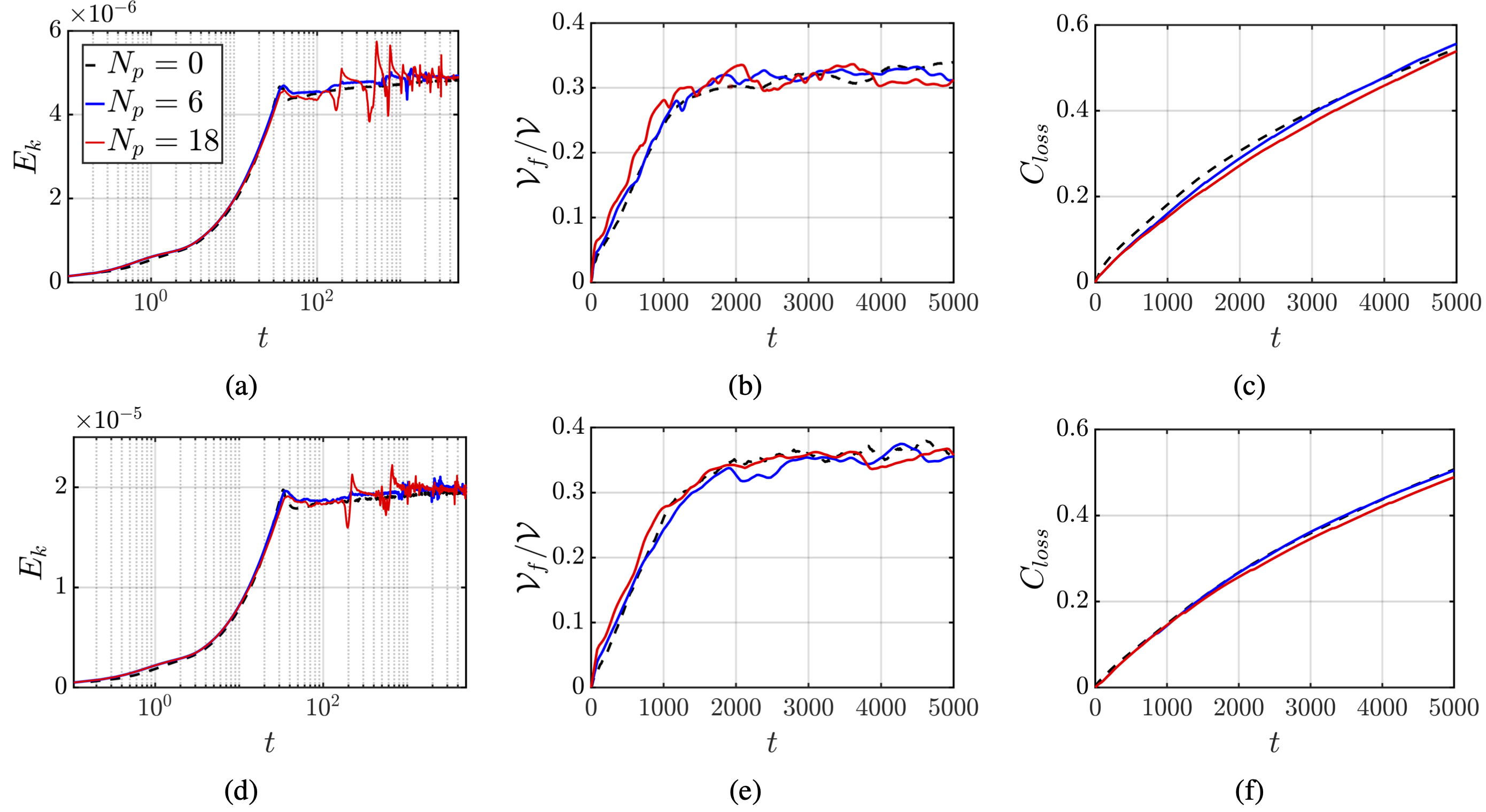}  
\caption{
The effect of rigid stones on the flow field as a function of the number, $N_p$, for $R_s=2$ and $Pe=500$. Kinetic energy, entrapped fluid volume when $Q>0$, and the clearance rate of a passive tracer are shown in columns one to three, respectively.
The top  and bottom panels correspond to  $\text{Re}=250$ and $\text{Re}=500$, respectively.
\label{concentration} }
\end{figure}

In Figure \ref{concentration}, we show the influence of stones on the fluid kinetic energy, defined as $E_k=\int_{\mathcal{V}} (\mathbf{u}^2/2)d\mathcal{V}$, the entrapped fluid volume for $R_s = 2$ \footnote{$R_s = 2$ was chosen as it is larger than the size of the outflow channel, $h$, and therefore the stones will remain within the cavity} and $Pe=500$, and the clearance rate of the passive tracer.
Inspection of the kinetic energy $E_k$ in figure \ref{concentration}a,d reveals a complex relationship with the number of stones, $N_p$. For all values of $N_p$, at early times ($t<35$), $E_k$ grows independently of the number of stones. Then the kinetic energy plateaus, but for $N_p>0$, we observe irregular oscillatory behaviour with increased amplitude for higher $N_p$. The fluid must do work to accelerate the stones, taking some of the kinetic energy of the fluid,  but since the motion of the stones is quite irregular, they experience numerous accelerations and decelerations. Hence, the more stones there are, the more noisy the kinetic energy.

%that increasing $N_p$ acts to increase the overall value of $E_k$. At early times ($t<35$)  $E_k$ grows independently of the presence of stones within the cavity as stones start stationary. At longer times, we observe the presence of oscillations in $E_k$ for $N_p>0$, with the amplitude of the oscillations increasing with the number of stones. This suggests that the motion of the rigid stones affects the flow field by accelerating and decelerating the fluid around the stones. 
%The amplitude of the oscillations are reduced when inertia is increased, and subsequently  the rigid solids have a stronger role in the flow dynamics at lower Reynolds numbers.

%

\begin{figure}[hbt!] 
\centering
\includegraphics[width=1\linewidth]{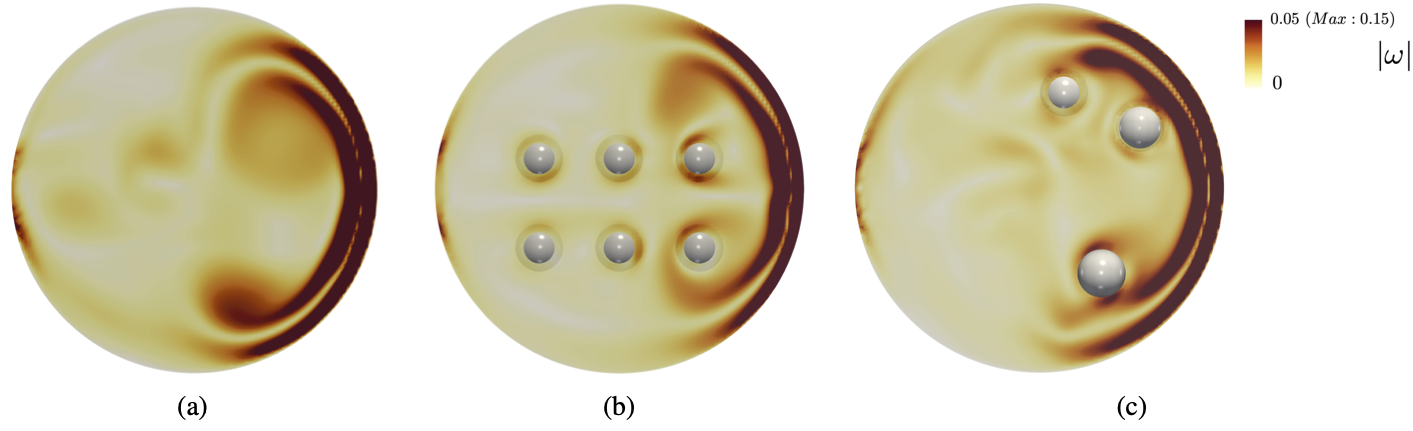}  
\caption{
Two-dimensional representation of the vorticity field  in the $xz$-plane ($y=2$) in the absence of stones, fixed stone location and stones which move freely in the cavity, corresponding to panels $(a-c)$, respectively. The conditions for the simulations are  $Re=250$ and $R_s=2$ and $N_p=18$ (for cases b and c).
The colour represents the  vorticity magnitude $|\omega|$, where appropriate scales are shown in the panel. Irregular vortex production can be seen around the rigid stones which move freely.
\label{vorticity_field} }
\end{figure}

Panels (b), (e) of figure \ref{concentration} plot the ratio of entrapped fluid volume (i.e. with $Q>0$) to total fluid volume. 
At early times, a linear growth of the entrapped fluid  volume is observed. As mentioned above there is a direct link between the entrapped volume and  vorticity production. \cite{batchelor_1967} showed that vorticity is produced at boundaries; thus, the addition of stones in the cavity implies the addition of regions for production of vorticity. However, the stones may also disrupt vortical structures, depending on their location within the cavity. This is demonstrated in figure \ref{vorticity_field}, which shows a snapshot of the instantaneous vorticity field (in a fixed plane) for three scenarios: (a) the absence of stones, (b) stones with position held fixed, and (c) freely moving stones. Comparing (b) and (c) to (a), we see the two-sided impact of stones, both in producing partially detached vortices, and in disrupting vortical structures; similar findings were reported by \cite{essmann_2020}. Due to this complex interplay, a cavity with more stones may have increased entrapped volume at some times, and decreased entrapped volume at other times, a noisy and unpredictable relationship as demonstrated in figure \ref{concentration} (b), (e).

%
%Although there is a reduction of the effective volume inside of the cavity based on the $N_p$, we observe that the entrapped fluid volume tends to similar values to the ones in the absence of stones, with regions with higher and lower entrapped fluid volume, suggesting that the presence of stones disrupts the vortical structures within the cavity.

We now turn our attention to the clearance rate plots (panels (c), (f) of figure \ref{concentration}), which show the percentage of tracer that has exited the cavity over time.
It has been previously demonstrated \cite{williams_2021} that the reduction of vortex structures within the cavity results in a reduction in the time needed for the clearance of the tracer, while clearance time depends non-monotonically on the fluid kinetic energy. Given the complex relationships between stone presence and both kinetic energy and vorticity, it is perhaps not surprising then that the clearance rate does not exhibit a clear trend. The highest number of stones does perform the worst in terms of tracer clear out, it is by a small margin, and the smaller number of stones performs slightly better than the base case of no stones.

\vspace*{-5pt}

\section{\textcolor{black}{Discussion}}

\textcolor{black}{A fundamental challenge in optimising protocols in ureteroscopy procedures is understanding the flow of irrigation fluid within the kidney, the movement of kidney stone particles and/or dust, and how larger stone fragments may impact fluid flow. Stone destruction via laser lithotripsy can create a range of sizes of stone fragments or dust \cite{aldoukhi2017holmium}, which, if not removed efficiently via irrigation fluid, can obscure the operating clinician's field-of-view \cite{moore2005minimally, smith2007smith}. Previous theoretical studies have uncovered the interplay between fluid structure, in particular the presence of vortical regions, and dust washout. These studies uncovered qualitative features and enabled large parameter exploration with minimal computational cost, but were restricted to a highly idealised two-dimensional geometry. To bridge the gap towards clinical relevance, in this paper we have extended this framework by incorporating discrete stone particles of finite size that both move with and modulate the flow, solved in a more realistic three-dimensional setting.}
%Time-dependent three-dimensional numerical simulations
The chosen parameters for the flow dynamics and the stone sizes are  consistent with clinically-realisable systems, and the numerical framework was validated against the experimental and numerical work presented by \cite{williams_2020} for the two-dimensional geometry.
%were carried out to study the flow dynamics inside of a renal pelvis in the context of kidney-stone removal. Particular attention was paid to the  temporal motion of spherical rigid-stones which move freely in the cavity. 
%. 
%

\textcolor{black}{In contrast to the aforementioned 2D studies, the vortex structure we have uncovered in the present study are more complex both in terms of spatial structure, interaction, and time-dependence. Specifically, we observed that}
the injection of a fluid jet via a modelled ureteroscope nozzle results in the formation of a primary vortex owing to the detachment of the velocity boundary layer from the nozzle.
The primary vortex grows over time, and its axisymmetric shape becomes more disrupted with increasing Reynolds number.
In the absence of stones, we have quantified the temporal entrapped fluid volume by identifying the recirculation zones  through the Q-criterion, i.e. $Q>0$, integrated over the cavity domain. At low inertia, a steady state is reached, and subsequently a constant volume for the entrapped-fluid  is predicted. At increasing inertia, a steady state is not reached owing to  the formation of a secondary vortex from the cavity walls. Mutual induction of the primary and secondary vortices characterise the complex flow dynamics observed in the system. 
The increase of fluid inertia at larger Reynolds numbers also results in an increase in the streamwise vorticity component with respect to its azimuthal vorticity counterpart. Thus, it is responsible for the loss of the axisymmetric behaviour observed in the small Reynolds number case.

With rigid stones introduced in the cavity flow, two different behaviours were observed: either they become trapped in a circulatory motion in the primary vortical structure or they are flushed out of the cavity together with the fluid through the ureteral access sheath. Both regimes were found to coexist in the studied cases and the outcome of the stones depended  on the initial location of the stationary stone, and the importance of inertia. More fluid inertia allows for higher probability of a stone to leave the cavity within the same temporal frame.
We have also studied the effect of multiple stones in the flow dynamics by exploring the fluid kinetic energy, entrapped fluid volume and the clearance rate of a passive tracer. 
We have shown that the solid boundaries, due to the the presence of stones, cause an increase in the  vorticity production, which results in richer flow dynamics within the cavity.

We have simulated for the first time (to the best of our knowledge) the two possible outcomes for the kidney stone trajectories during ureteroscopy -- trapping or wash-out -- and this can be considered to be a significant step forward in the understanding of the dynamics of kidney stones in the renal cavity flow during the surgical procedure. \textcolor{black}{The complex behaviour outlined above highlights the need for computational fluid mechanics tools in ureteroscopy, and demonstrates a strong potential for optimising driving flow conditions to promote stone removal. In practice, this may best be achieved with time-dependent driving flow, a complication we have not considered here.}

\begin{figure}%[hbt!] 
\centering
\includegraphics[width=1\linewidth]{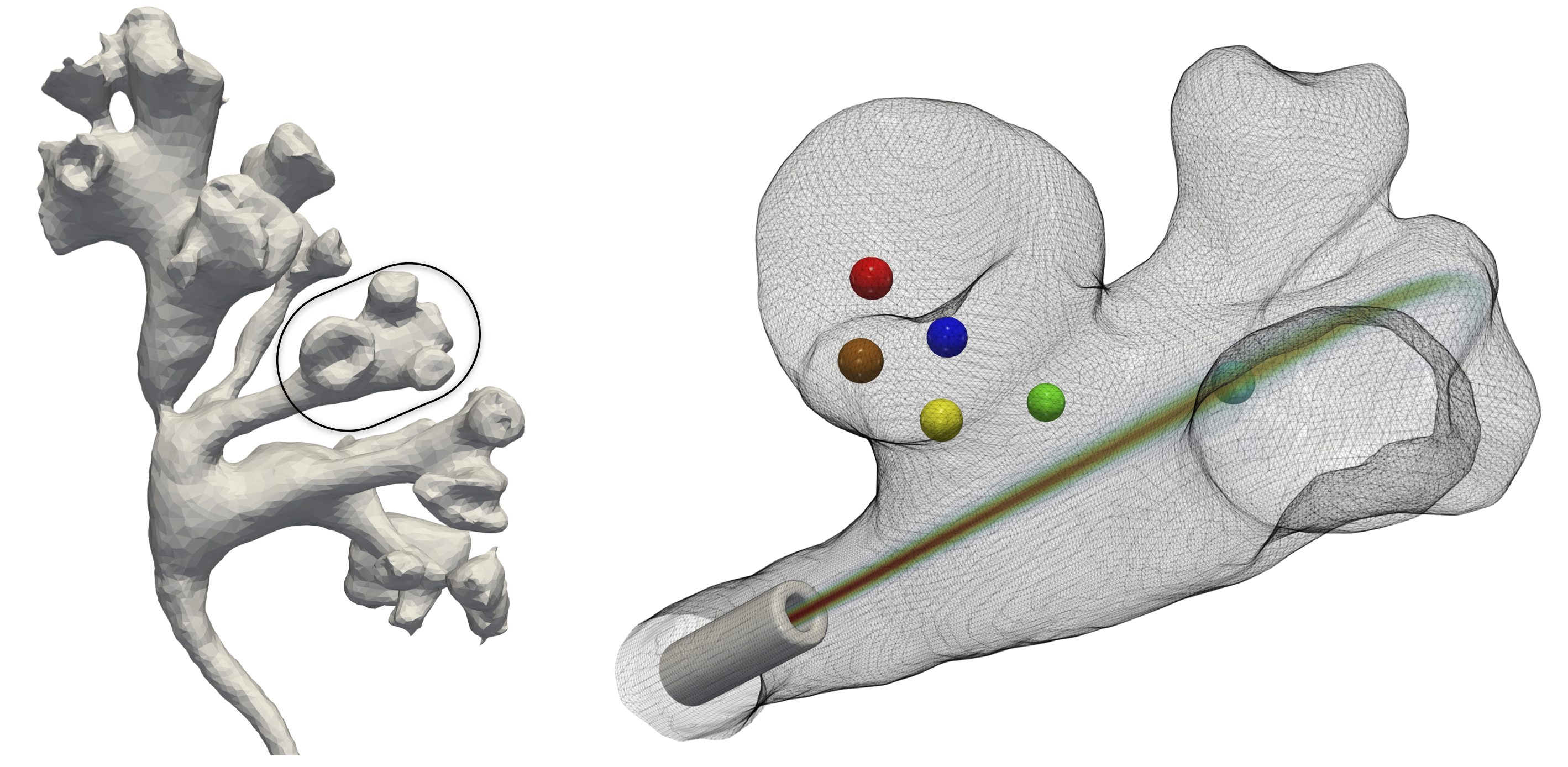}  
\caption{ The left panel shows a three-dimensional renal pelvis  from a patient which has been reconstructed using CT scans. The right panel shows a snapshot at long times for a preliminary simulation  to demonstrate the capabilities of the methodology used in this research. We have extracted one of the branches of the renal pelvis (highlighted in the left panel), and have introduced a ureteroscope. \textcolor{black}{The close interaction of flow cavity and the stiff stones result  in the stones being trapped in one of the renal pelvis branches.} 
 Computed velocity is shown to visualise the flow dynamics inside of the cavity.
\label{perspectives} }
\end{figure}

\textcolor{black}{As well as incorporating time-dependent inlet conditions, t}he approach we described in this manuscript may naturally be extended to a  systematic follow-up analysis predicting the flow characteristics and stone motions in an actual renal cavity geometry, potentially incorporating full three-dimensional time-dependent simulations and coupling of moving solids, buoyancy effects, and heat-transfer considerations.  \textcolor{black}{We have restricted attention to an idealised spherical cavity, and an actual renal pelvis would provide a more representative interaction of stones with the fluid}. Figure \ref{perspectives} shows a preliminary simulation from an actual renal pelvis extracted from imaging in a patient, in which we have selected a branch  of the complex renal pelvis and introduced the ureteroscope \textcolor{black}{ together with rigid stones; results at long times show that the stones are trapped in one of the branches  of the renal cavity}. 
This avenue of research could lead to the next generation of in-silico models that are patient-specific, and translate our understanding ‘from bench-to-bedside’.

The present work is limited by the assumption of rigid cavity walls. This assumption is realistic for the inflow and outflow  channels; however, the walls of an actual renal cavity are characterised by deformability that is inherent to any true physiological structure. Deformability of the renal pelvis would certainly affect the flow structure inside of the cavity, potentially resulting in more complex flow dynamics, and constituting a fruitful area of future research. The numerical method used in this study may naturally be extended to include deformability of the renal cavity and inclusion of these effects constitutes a realistic short-term follow-up to the present study.
\textcolor{black}{We have limited our study to perfectly spherical kidney stones, but their inherent complex three-dimensional  shape will play a major role in the fluid-structure interaction. We have only studied equal-sized rigid stones in the cavity, but in surgical procedures, stone destruction via laser lithotripsy leads to obliteration of the stone into a large range  of sizes, and subsequently, future research should consider the potential effects of uneven-sized stones. Particle-particle interaction should be taken into account in future studies as it would become more important as the particles become larger and more numerous.} \textcolor{black}{Additionally, future studies  would also quantify the enstrophy within the renal cavity as a function of the time, as a measure of the rotational energy of the fluid flow.}

\section{\textcolor{black}{Conclusions}}

\textcolor{black}{This work has focused on understanding the fluid mechanics of ureteroscope irrigation through three-dimensional numerical simulations. This study is  limited to stiff, non-deformable solids for both the stones and the renal pelvis cavity. The numerical predictions show the close interplay between the local vorticity and the stones, and their interaction decides the  outcome of the stones: either they become trapped in a circulatory motion in the primary vortical structure or they are flushed out of the cavity together with the fluid through the ureteral access sheath. We have also studied the efficacy of debris clearance influenced by the presence of different numbers of  kidney stones.}

\section*{Acknowledgements}
We  acknowledge HPC facilities and computing resources provided by the Research Computing Service (RCS) of Imperial College London and support through computing time at the Institut du Developpement et des Ressources en Informatique Scientifique (IDRIS) of the Centre National de la Recherche Scientifique (CNRS), coordinated by GENCI (Grand Equipement National de Calcul Intensif) Grant No. 2022A0122B06721. L.K. acknowledges the financial support from the Engineering and Physical Sciences Research Council, United Kingdom, through the EPSRC PREMIERE (EP/T000414/1) Programme Grant. The numerical simulations were performed with code BLUE (\cite{Shin_jmst_2017}) and the visualisations have been generated using ParaView.
\\
\\

The authors declare no conflict of interest.

\bibliographystyle{asmems4}
\bibliography{sample}

%%%%%%%%%%%%%%%%%%%%%%%%%%%%%%%%%%%%%%%%%%%%%%%%%%%%%%%%%%%%%%%%%%%%%%
\appendix       %%% starting appendix
\section*{Appendix: Validation of the numerical method}

This section provides some validation studies for the immersed solid solver, flow-solver and mesh studies to provide conclusive evidence of the accuracy of our numerical predictions.  

% \subsubsection*{Validation of the numerical method}

\begin{figure}
\centering
\includegraphics[width=1\linewidth]{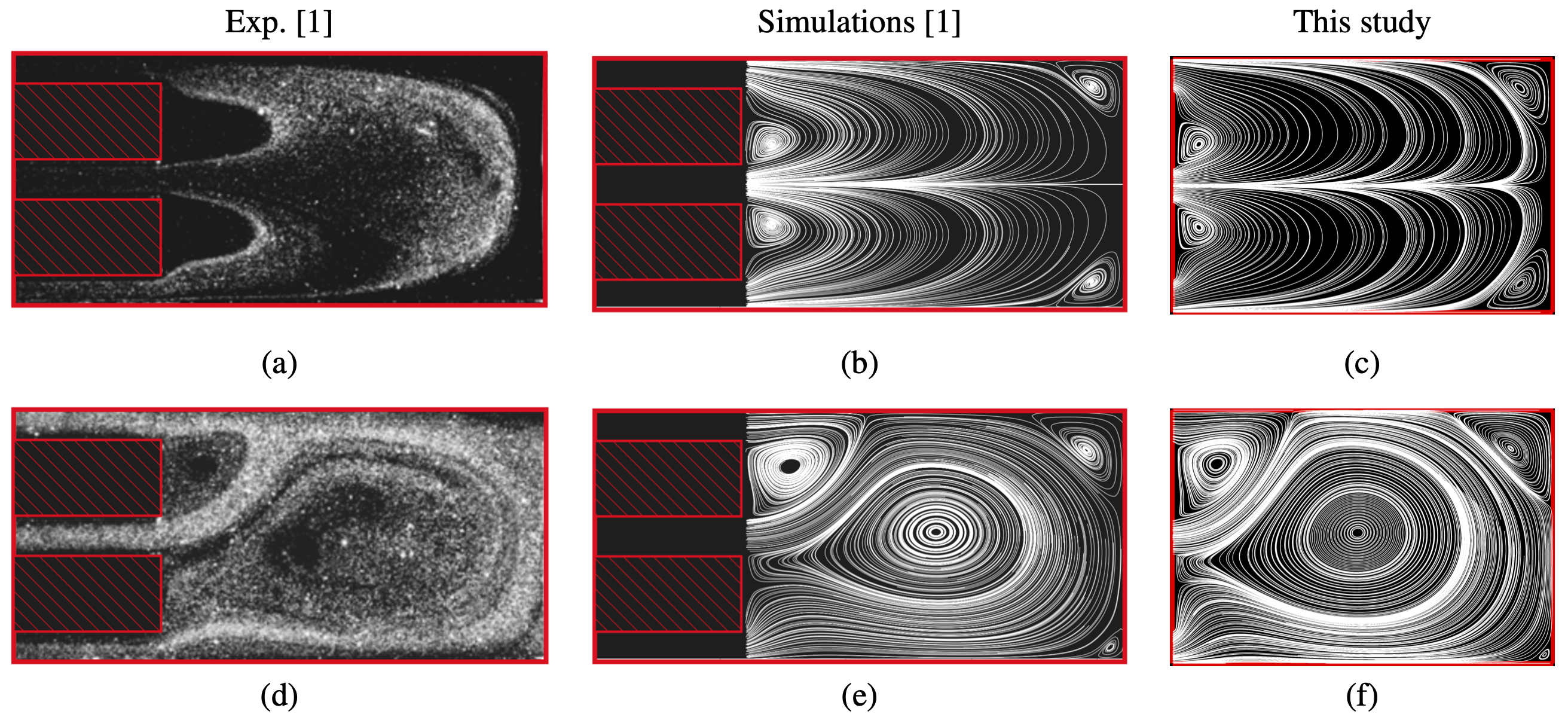}  
\caption{ Direct comparison of the streamlines from  our numerical predictions, third column, with both experimental and simulation results reported by \cite{williams_2020}, first and second columns, for a two-dimensional cavity characterised by $b = 1.933$, $h = 4.2835 $, and cavity-length $\alpha=21.7$ (here, $\alpha$ stands for the nondimensional length of the rectangular cavity). Panels (a)-(c) and (d)-(f) represent the flow dynamics with  $\text{Re}=7$ and $\text{Re}=34$, respectively.
\label{validation} }
\end{figure}

In order to assess the accuracy of our flow solver, we have validated our predictions against the experimental and numerical data from \cite{williams_2020}.
Figure \ref{validation} highlights qualitative validation of our numerical frame-work 
in terms of  numerical streamlines for $\text{Re}=7$ and $\text{Re}=34$.  
It is clear that the transient solutions of the Navier Stokes equations provided by our numerical method are capable of predicting the rich dynamics observed by the flow-visualisation experiments, and previous numerical simulations based on the steady solutions of the flow equations. Additionally, we want to highlight that the flow-solver has also been successfully validated for two-phase turbulent jets in previous works, such as  \cite{Constante_gfm,constante_jets,phd_thesis_CRCA}.

\begin{figure}
\centering
\includegraphics[width=1\linewidth]{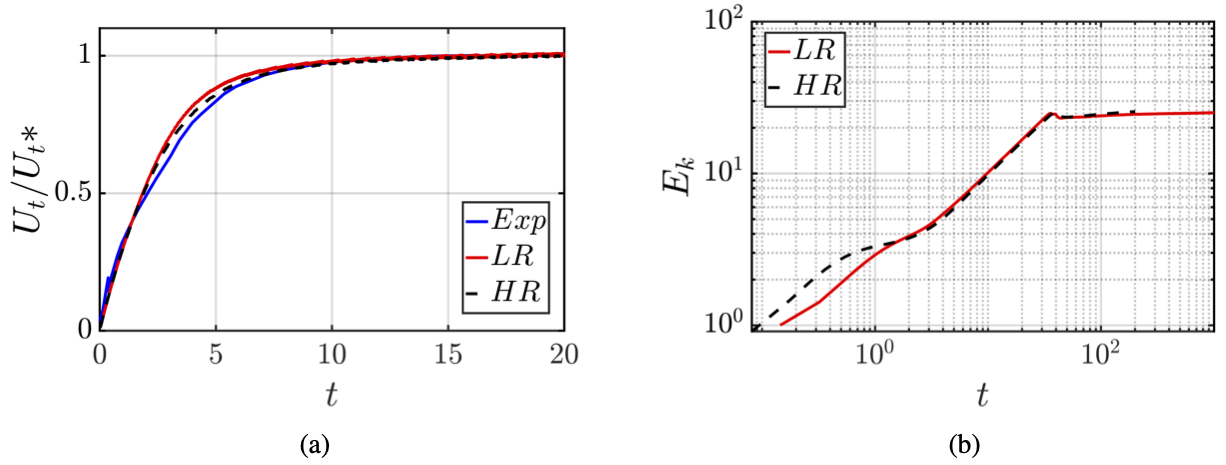}  
\caption{ 
(a) Settling velocity of a solid sphere at $Re=41$ in relation to the experimental results reported by \cite{Mordant}, here the velocity has been normalised with respect to the experimental observation at steady state, $U_t*$. 
(b) Mesh study for $Re=250$ in the absence of stones highlighting the  temporal evolution of the kinetic energy $E_k$. 
\label{validation_solid} }
\end{figure}

For the  validation of  the immersed solid solver using the Direct Forcing approach of \cite{Fadlun}, we have considered the transient and terminal velocities of a solid sphere settling under gravity in a quiescent fluid. We compare our numerical predictions   against the experiments of
\cite{Mordant}, where  solid spheres are released in water. 
We have considered the case of a particle with density ratio of $\rho_s/\rho_l=2.56$ and $Re= d_s \rho_s U_t/ \mu_l=41$, where$ d_s$ and $U_t$ correspond to the particle diameter and terminal velocity of the particle (e.g. computations are assumed to reach steady-state when the settling velocity is below $0.1\%$ for one period of dimensionless time).
The experiments were done under the assumption of an unbounded liquid bath, therefore our simulations are performed in  a sufficiently large domain to avoid effects from the boundaries. 
Figure \ref{validation_solid}a shows the predicted settling velocity  for two different levels of refinement (e.g., $6$ and $10$ cells per sphere radius, which corresponds to the low (`LR') and high resolution (`HR')  simulations, respectively). 
As shown in figure \ref{validation_solid}a, the predictions from the immersed solid solver present an excellent   agreement with respect to   experimental measures for the transient and  terminal velocity (e.g., the latter with error of less than  $0.1\%$).
Additionally, we acknowledge that the same numerical method has been previously validated with respect to the interaction of a moving solid with the flow, we refer the reader to \cite{Shin_2020}, who described extensively the numerical formulation and provided extensive benchmark tests of the fluid-structure interaction of  solids (either as rigid or deformable structures) with multiphase  flows using a combination of immersed boundary and fictitious domain-direct forcing methods. 

Finally, we aim to provide conclusive evidence that  our numerical results are mesh-independent. To this end, the kinetic energy for $Re=250$ in the absence of stones are tested for different mesh resolutions.
Figure \ref{validation_solid}b shows the temporal evolution of $E_k$ for two types of refinements (e.g., `LR' and `HR' refer to levels of refinement characterised by $a/\Delta {\bf x}=6.4$ and $a/\Delta {\bf x}=12.8$, respectively).
As shown in figure \ref{validation_solid}b, both level of refinements are capable of predicting accurately the dynamics  at steady state. Thus,  we conclude that a mesh characterised with $a/\Delta {\bf x}=6.4$ is sufficiently refined to ensure mesh-independent results while providing a good compromise with the computational cost of the simulation.
Therefore, we have proved that the `LR' mesh is capable of predicting the complex dynamics of the phenomena, and consequently detailed analysis of the vortical
structures is performed using a `LR' mesh-type (unless stated otherwise).

\end{document}